\documentclass[aps,prl,twocolumn,footinbib,showpacs,floatfix]{revtex4}
\usepackage{amssymb}
\usepackage[pdftex]{graphicx}	
	\DeclareGraphicsExtensions{.pdf,.png,.jpg}
\usepackage[applemac]{inputenc}
\usepackage{amsmath,amssymb,amsthm}
\usepackage{mathbbol,bbm}
\usepackage{graphicx,float}
\usepackage[final]{showkeys}
\usepackage{bm,dsfont}
\usepackage{color}
\usepackage{hyperref}

\definecolor{myblue}{rgb}{0.3,0.4,0.85}
\definecolor{myred}{rgb}{0.8,0.,0.2}
\definecolor{mygreen}{rgb}{0.6,0.8,0.2}

\def\ket#1{\left|#1\right>}
\def\bra#1{\left<#1\right|}

 \def\ii{\mathord{\rm i}}

\renewcommand{\ii}{{\rm i}}

\usepackage{booktabs}

\begin{document}
\title{Topological Massive Dirac Edge Modes \\
and Long-Range Superconducting Hamiltonians}
\author{O. Viyuela$^1$, D. Vodola$^2$, G. Pupillo$^2$ and M.A. Martin-Delgado$^1$ }
\affiliation{1. Departamento de F\'{\i}sica Te\'orica I, Universidad Complutense, 28040 Madrid, Spain.\\
2. icFRC, IPCMS (UMR 7504) and ISIS (UMR 7006), Universit\'e de Strasbourg and CNRS, 67000 Strasbourg, France.}

\vspace{-3.5cm}

\begin{abstract}
We discover novel topological effects in the one-dimensional Kitaev chain modified by long-range Hamiltonian deformations in the hopping and pairing terms.
This class of models display symmetry-protected topological order measured by the Berry/Zak phase of the lower band eigenvector and the winding number of the Hamiltonians.
For exponentially-decaying hopping amplitudes, the topological sector can be significantly augmented as the penetration length increases, something experimentally
achievable. For power-law decaying superconducting pairings, the massless Majorana modes at the edges get paired together into a massive non-local Dirac fermion localised at both edges of the chain: a new topological quasiparticle that we call \emph{topological massive Dirac fermion}.
This topological phase has fractional topological numbers as a consequence of the long-range couplings.
Possible applications to current experimental setups and topological quantum computation are also discussed.
\end{abstract}


\maketitle


\noindent {\it 1. Introduction.---}  The quest for the experimental realisation of topological superconductors
has turned out to be far more elusive than for their insulating counterparts.
Simple models for topological superconductors have been proposed \cite{rmp1,rmp2}, but yet their unambiguous implementation is challenging in condensed matter or with quantum simulations. Here we address the issue as to whether those simple models \cite{Read_et_al00,Kitaev01} are in fact very specific in hosting their long sought-after topological properties. Quite on the contrary, we find that these properties can not only be generic with respect to natural extensions of the model-Hamiltonian terms, but also that Hamiltonian deformations can give rise to unconventional topological edge-mode physics that is novel per se and for applications in topological quantum computation.

The appearance of topological superconductors is having a strong impact \cite{Mourik_et_al12,Deng_et_al12,Das_et_al12,Nadj_et_al14,Sun_et_al16,Albrecht_et_al16} in condensed matter physics and quantum simulators. 
A tremendous effort is now directed at the experimental demonstration of existing topological models and at the development of new ones that may be easier to realise.
What makes a topological superconductor interesting is the presence of Majorana modes as zero-energy localized modes
at the edges or boundaries of the material. These modes lie within the superconducting gap and are rather exotic since Majorana
fermions are their own anti-particles (holes). Standard (non-topological) superconductors do not exhibit such modes
in their energy spectrum. Thus, topological superconductors represent new physics: Majorana modes are topologically
protected against local perturbations disturbing the system and cannot be removed unless a topological phase transition occurs.
This robustness makes them useful for storing and manipulating quantum information in a topological quantum computer.

\begin{figure}[t]
\centering
\includegraphics[width=\columnwidth]{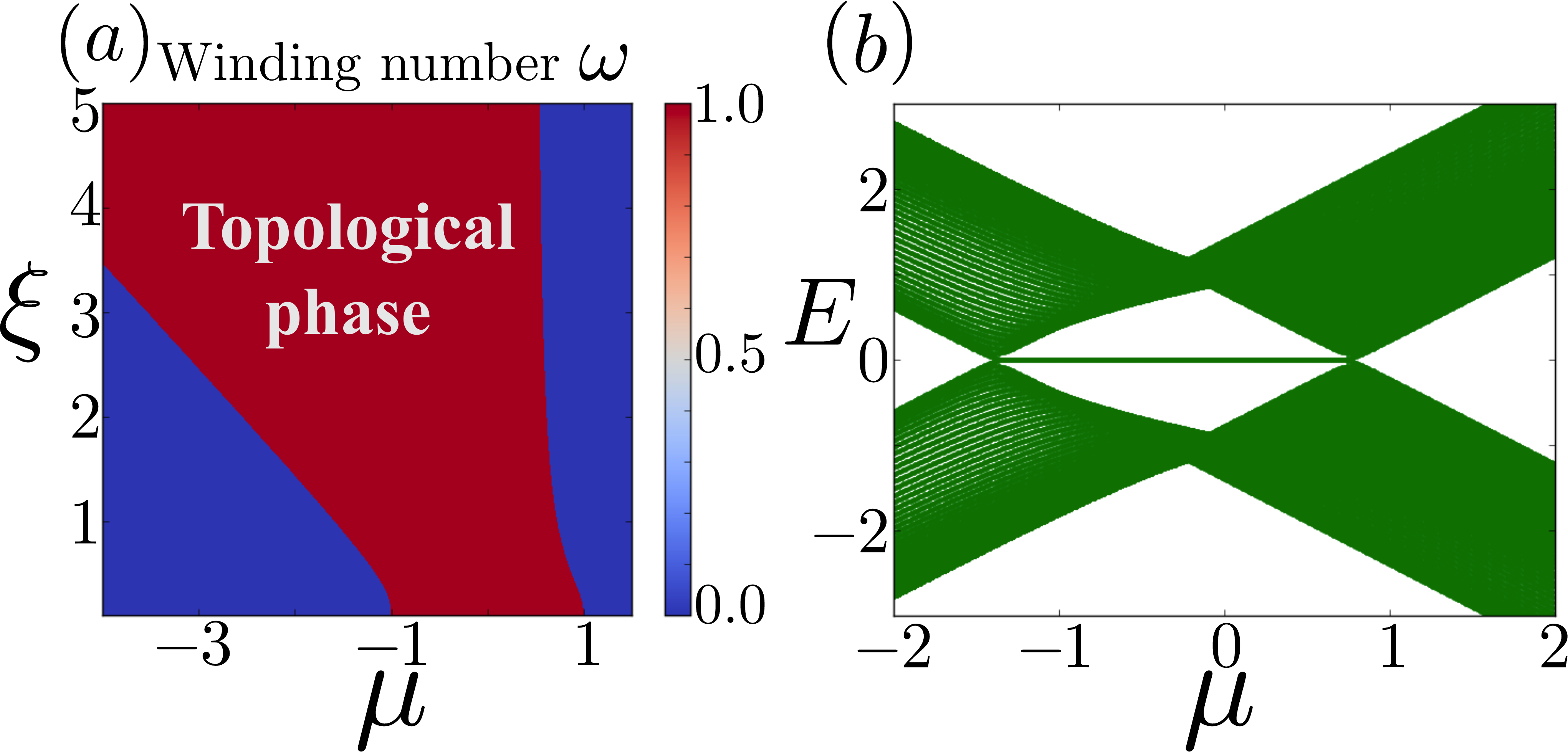}
\caption{(a) Topological phase diagram for the Kitaev chain with exponentially decaying hopping. As the penetration length $\xi$ increases, the topological phase ($\Phi_{\rm B}=\pi\omega=\pi$) gets enlarged. For $\xi\rightarrow0$ we recover the well-known Majorana chain with nearest-neighbour hopping only. (b) Energy spectrum for $\xi=0.8$. The region with MZMs $\mu \in (-1,1)$ in the original model has been augmented in one to one correspondence with a non-trivial Berry phase and winding number.} 
\label{exp_phase}
\end{figure}

\begin{figure}[t]
\centering
\includegraphics[width=\columnwidth]{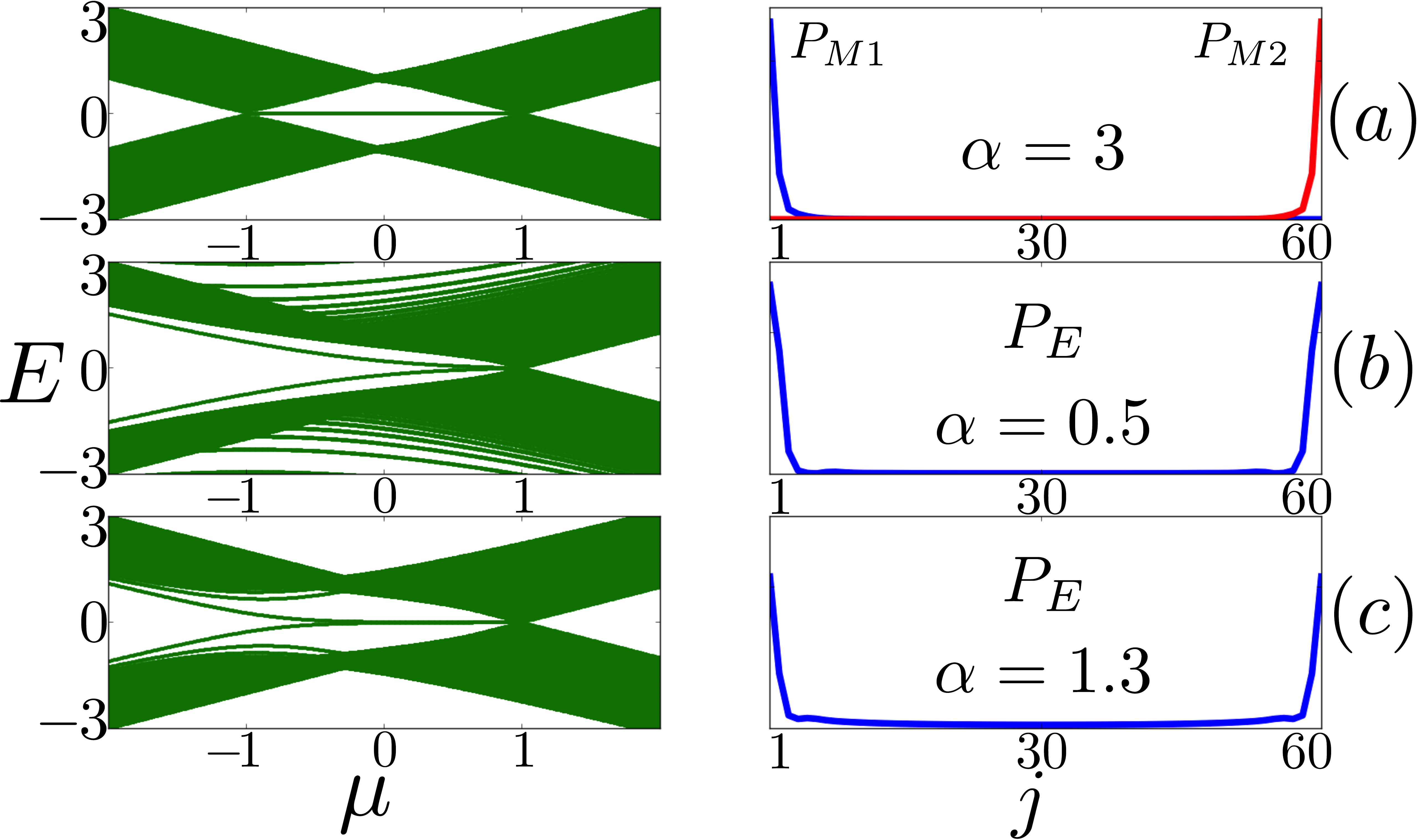}
\caption{Left side, we plot the spectrum for the Kitaev chain with long-range decaying pairing with $R_{l,\alpha}$, for $L=60$ sites. On the right hand side we show the probability distribution $P_E$ of the edge modes for different topological phases. (a) Majorana sector with $\alpha=3$. We can see MZMs for $\mu\in[-1,1]$ localised at the edges of the chain, as plotted on the right hand side for $\mu=-0.5$ ($P_{M1}$ and $P_{M2}$). Notice that each Majorana mode is decoupled, represented with different colors. (b) Massive Dirac sector with $\alpha=0.5$. Within the new topological phase ($\mu<1$), there are topological massive Dirac fermions localised at both edges at the same time, as shown on the right hand side for $\mu=-1.5$. (c) Crossover sector with $\alpha=1.3$. There are both MZMs and massive Dirac fermions depending on the value of $\mu$. We plot the probability for a massive Dirac fermion at $\mu=-1.2$.}
\label{Top_Dirac}
\end{figure}

\begin{figure}[t]
\centering
\includegraphics[width=\columnwidth]{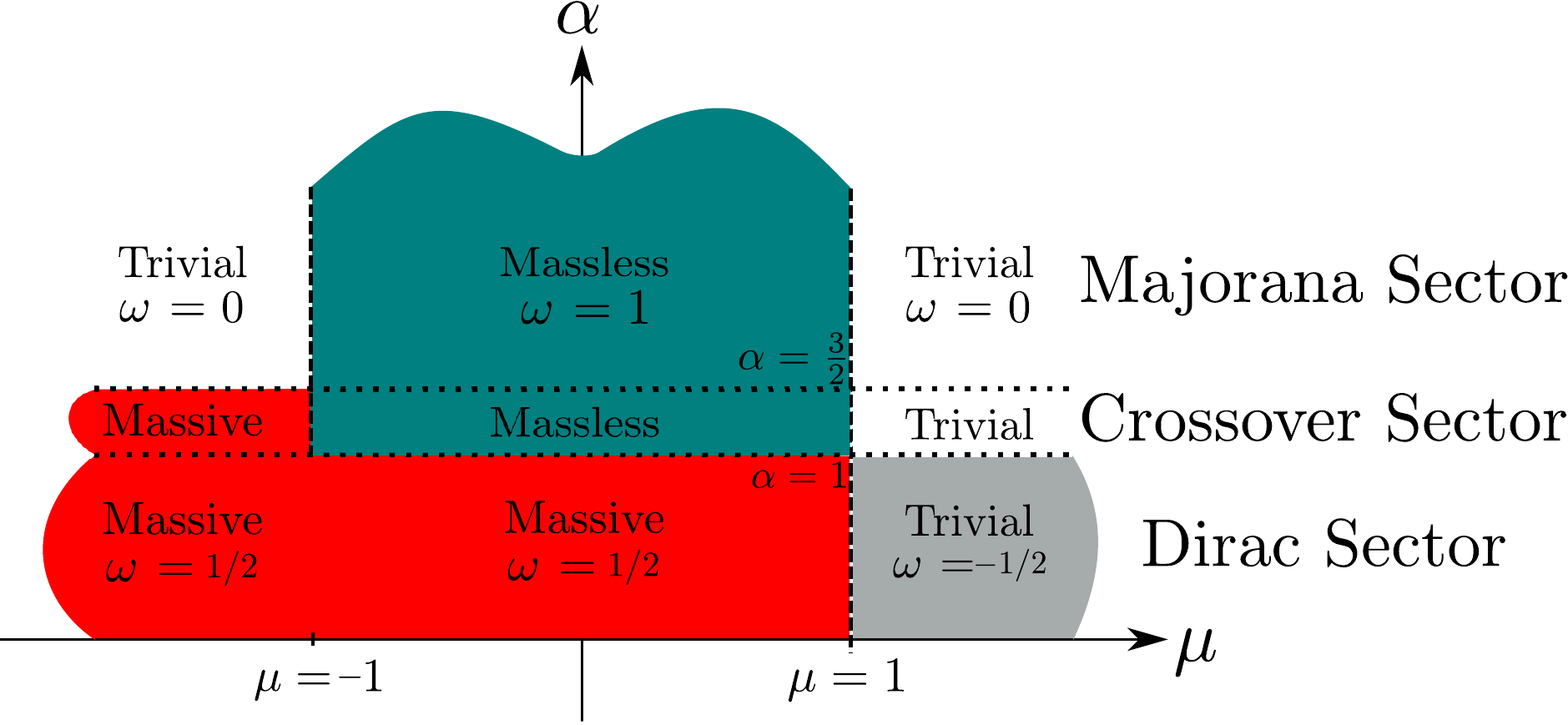}
\caption{Topological phase diagram for the Kitaev chain with long-range pairing. The wavy lines at the border of certain phases indicate that they extend endlessly. Fractional topological numbers highlight the appearance of an unconventional topological phase with massive non-local Dirac edge states. The topological characterisation of the crossover sector is discussed in the main text and the Appendix.} 
\label{pairing_phase}
\end{figure}

In this paper we focus on the Kitaev chain model and propose novel modifications
of the basic Hamiltonian, in order to enrich the appearance of Majorana physics (see Fig.~\ref{exp_phase}) and even new topological excitations
(see Fig.~\ref{Top_Dirac}, Fig.~\ref{pairing_phase}). These modifications come in two ways: a) exponentially decaying kinetic terms and b) long-range (LR) interaction terms. They produce novel beneficial topological effects and  new unconventional topological physics, respectively.
In case a), we propose a hopping deformation
that allows us to significantly increase the region in the phase diagram where
Majorana zero modes (MZMs) are present. Interestingly enough, this modification
may result in a realistic description for cold atoms in optical lattices.
In case b), we study the topological properties of another complementary modification of the
Kitaev model based on long-range pairing terms decaying algebraically with
a certain exponent $\alpha$. We discover novel topological effects not found in any
simple model before (see Fig.~\ref{Top_Dirac}):
 for $\alpha < 1$ the model suffers a major qualitative change manifested in the
absence of MZMs that are transmuted onto Dirac modes, which
are massive non-local edge states. These new edge states are topologically protected against perturbations that do not break fermion parity nor particle-hole symmetry. These modes appear as mid-gap superconducting states that cannot be absorbed into bulk states. These topological massive Dirac
edge states are new physical quasiparticles that are absent in the standard
Kitaev model. They represent a new unconventional topological phase.


\noindent {\it 2. Long-range deformations of superconducting Hamiltonians.---}
We consider a model of spinless fermions on a $L$-site one-dimensional chain, with {\it p}-wave superconducting pairing and a hopping term. The Hamiltonian of the system is
\begin{eqnarray}
H&=&\sum^L_{j=1}\Big(-J\sum_{l=1}^{L-1}\frac{1}{r_{l,\xi}}a^{\dagger}_ja_{j+l}+M\sum_{l=1}^{L-1}\frac{1}{R_{l\alpha}}a_ja_{j+l}-\nonumber\\
&-&\frac{\mu}{2}(a_j^{\dagger}a_j-\frac{1}{2})+\text{h.c}\Big),
\label{H}
\end{eqnarray}
where $\mu$ is the chemical potential, $J>0$ is the hopping amplitude, the absolute value of $M=|M|\text{e}^{{\rm i}\Theta}$ stands for the superconducting gap, $a_j$ $(a_j^{\dagger})$ are annihilation (creation) fermionic operators. The Hamiltonian deformations are $r_{l,\xi}$, $R_{l,\alpha}$. They are generic functions of an integer distance $l$, and parameters $\xi$ and $\alpha$, respectively. The total number of fermions modulo 2 is called the 'fermion parity' and it is a conserved quantity for all models in \eqref{H}. 
Considering only nearest-neighbours hopping and pairing, we recover the famous model introduced by Kitaev \cite{Kitaev01}. This model is topological displaying MZMs at the edges like in Fig.~\ref{Top_Dirac}(a). 
In the topological phase, the ground state of the Kitaev model is two-fold degenerate: a bulk of with even fermion parity, while populating the two Majorana modes at the edges amounts to a single ordinary fermion and odd parity. The conservation of fermion parity and the non-local character of the unpaired Majoranas at the edges make the system an ideal candidate for a topological qubit out of the two-fold degenerate ground state \cite{rmp3,rmp4}.

Without loss of generality, we may fix the pairing amplitude to be real and $M=J=\frac{1}{2}$. Assuming periodic boundary conditions, 
we can diagonalize the Hamiltonian deformations \eqref{H} in Fourier space and in the Nambu-spinor basis 
representing  paired fermions \cite{ASBook}:  $H=\frac{1}{2}\sum_k\Psi_k^{\dagger}H_k\Psi_k$, where $\Psi_k=\big(a_k,a^{\dagger}_{-k}\big)^{\rm t}$ and $H_k$ is of the form $H_k={E_k}{\bm n}_k\cdot{\bm \sigma}$. The energy dispersion relation is given by $E_k$, ${\bm \sigma}$ is the Pauli matrix vector and ${\bm n}_k$ is a unit vector called \emph{winding vector}. Explicitly,
\begin{eqnarray}
{\bm n}_k&=&-\frac{1}{E_k}\Big(0,f_{\alpha}(k),\mu+g_{\beta}(k)\Big),\nonumber\\		
E_k&=&\sqrt{(\mu+g_{\xi}(k))^2+f_{\alpha}^2(k)},
\label{nMC}
\end{eqnarray}
with 
\begin{equation}
g_{\xi}(k)=\sum_{l=1}^{L-1}\frac{\cos{(k\cdot l)}}{r_{l,\xi}}~~\text{and}~~f_{\alpha}(k)=\sum_{l=1}^{L-1}\frac{\sin{(k\cdot l)}}{R_{l,\alpha}}. 
\label{gf}
\end{equation}

Particular instances of the functions $r_{l,\xi}$ and $R_{l,\alpha}$ have been considered in \cite{Vodola_et_al14,Vodola_et_al15}, where long-range deformations of the Kitaev chain were first considered. 

These models \eqref{nMC} belong to the BDI symmetry class of topological insulators and superconductors \cite{Ludwig,Kitaev_2009}, with particle-hole, time-reversal and chiral symmetry. The inclusion of long-range effects do not break these symmetries, nor the conservation of fermion parity. This is an important condition for the topological character of the original short-range model to be preserved. These symmetries impose a restriction on the movement of the winding vector ${\bm n}_k$ from the sphere $S^2$ to the circle $S^1$ on the 
$yz-$plane. Thus, we have a mapping from the reduced Hamiltonians $H_k$ on the Brillouin zone $k\in S^1$ onto the 
winding vectors ${\bm n}_k \in S^1$. This mapping 
$\mathit{S}^1\longrightarrow \mathit{S}^1$ is characterized by a winding number $\omega$, a topological invariant defined as the angle swept by ${\bm n}_k$ when the crystalline momentum $k$ is varied across the whole Brillouin Zone (BZ) from $-\pi$ to $+\pi$, 
\begin{equation}
\omega:=\frac{1}{2\pi}\oint d\theta=\frac{1}{2\pi}\oint\bigg(\frac{\partial_k {n}_k^z}{n_k^y}\bigg)dk,
\label{w}
\end{equation}
where we have used that $\theta:=\arctan\big({n}^z_k/{n}^y_k\big)$.

As a complementary tool in 1D systems, we can use the Berry/Zak phase \cite{Berry84,Simon85,Zak89} to characterize topological order. When the system is adiabatically transported from a certain crystalline momentum $k_0$ up to $k_0+G$, where $G$ is a reciprocal lattice vector, the eigenstate of the lower band of the system $\ket{u^{-}_k}$ picks up a topological Berry phase given by
\begin{equation}
\Phi_{\rm B}=\oint A_{\rm B}(k) dk.
\label{phiB}
\end{equation}
The Berry connection $A_{\rm B}(k)={\rm i}\bra{u^{-}_k}\partial_k u^{-}_k\rangle$ connects by means of a parallel transport two infinitesimally close points on the manyfold defined by $\ket{u^{-}_k}$ in $k$ space. For the standard Kitaev chain~\cite{Kitaev01}, the resulting gauge-invariant phase $\Phi_B$ is quantized ($0$ or $\pi$) due to the particle-hole symmetry that characterises  distinct topological phases in one to one correspondence with the winding number \cite{Viyuela_et_al14}.

\noindent {\it 3. Augmented topological phases induced by exponentially decaying hoppings.---}
This remarkable effect is obtained choosing nearest-neighbour pairing, i.e., $R_{1,\alpha}=1$, $R_{l>1,\alpha}=\infty$ and $r_{l,\xi}={\rm e}^{\frac{(l-1)}{\xi}}$, where  $\xi$ is the penetration length of the exponentially decaying hopping terms. This Hamiltonian may be realisable in simulations of topological superconductors using cold atoms in optical lattices \cite{Lewenstein10,Jian_et_al11,Buhler10}, where the exponential decay of the hopping terms with distance can be tuned, e.g., by varying the depth of the lattice potentials~ \cite{rmp5}.

In Fig.~\ref{exp_phase} we plot the complete phase diagram by computing the winding number and the topological Berry phase from Eqs.~\eqref{w} and \eqref{phiB}. For $\xi\rightarrow0$ we recover the usual  Kitaev chain. The system is topological for $\mu\in[-1,1]$, displaying MZMs at the edges. Interestingly enough, when we increase the penetration length $\xi$, the region where we observe MZMs is augmented. In fact, this widening effect is purely due to the hopping deformation since we find
that including an exponentially-decaying pairing deformation does not change the topological phases. 
In the thermodynamic limit $L\to\infty$, the phase separation between the trivial and non-trivial topological phases can be computed analytically from Eq.\eqref{nMC}, obtaining
\begin{equation}
\mu_{c1}=\frac{{\rm e}^{\frac{1}{\xi}}}{1+{\rm e}^{\frac{1}{\xi}}},~~\mu_{c2}=\frac{{\rm e}^{\frac{1}{\xi}}}{1-{\rm e}^{\frac{1}{\xi}}}.
\label{muc}
\end{equation}
Thus, increasing the penetration length of the deformed hopping, we can arbitrarily enlarge the topologically non-trivial sector (see Fig.~\ref{exp_phase}). 
Although symmetry-protected topological order is usually associated with local interactions, we have shown that non-local terms can favour the formation of a topological phase. Related studies for the Kitaev chain with long-distance hopping were carried out~\cite{DeGottardi_13} and qualitatively similar effects have been recently observed in Ref.~\cite{Gong2015_2} for the spin-1 long-range Haldane model~\cite{Gorshkov15}.

\noindent {\it 4. Unconventional topological superconductivity with Dirac topological massive states---}
Long-range deformations may not only enlarge topological phases but also produce new types of topological phases.
To this end, let us now consider pairing terms that decay algebraically with a power-law exponent $\alpha$, and no deformation of the hopping terms. That is, $r_{1,\xi}=1$, $r_{l>1,\xi}=\infty$ and $R_{\alpha,l}= l^{\alpha}$.

\noindent In the thermodynamic limit $L\rightarrow\infty$, the function $f_{\alpha}(k)$ in Eq.~\eqref{gf} is divergent at $k=0$ for $\alpha<1$. This function defines the long-range pairing and appears in the energy dispersion relation and the winding vector of Eq.~\eqref{nMC}. Thus, the dispersion relation and the group velocity also become divergent at $k=0$  if $\alpha<1$.
Nevertheless, $\omega$ [Eq.~\eqref{w}] and $\Phi_{\rm B}$  [Eq.~\eqref{phiB}] are still integrable. Moreover, it is not possible to gauge away the divergence from $k=0$ by means of a gauge transformation, as in the ordinary Kitaev chain. Therefore, the 
divergence behaves as a \emph{topological singularity}. A detailed discussion of this effect at $k=0$ on the topological indicators is carried out in Sec.~I  of the Appendix.
According to the behaviour of $f_{\alpha}(k)$ at $k=0$, we find 3 different topological sectors depending on the exponent $\alpha$:

i/ Majorana Sector [$\alpha>{3}/{2}$]--- this sector is topologically equivalent to the one of the short-range Kitaev chain \cite{Kitaev01}: For $\lvert{\mu}\rvert>1$, the phase is topologically trivial and we do not find MZMs. In the region $\mu\in(-1,1)$, we find that MZMs are always present [see Fig.~\ref{Top_Dirac}(a)]. The function $f_{\alpha}(k)$ is not divergent and we can compute the winding number $\omega$ of Eq.~\eqref{w} and the Berry phase $\Phi_{\rm B}$ of Eq.~\eqref{phiB} obtaining $\Phi_B=\pi\omega=\pi$. The lower band eigenvector $\ket{u^{-}_{k}}$, thus, shows a $U(1)$ phase discontinuity at $k=0$.
The corresponding topological phase is depicted in blue in the phase diagram of Fig.~\ref{pairing_phase}.

ii/ Massive Dirac Sector [$\alpha<1$]--- an unconventional topological phase appears for sufficiently slow decaying pairing. As an example, in Fig.~\ref{Top_Dirac}(b)  we see for $\alpha=1/2$ two clearly different phases as a function of $\mu$. For $\mu>1$ the system is in a trivial phase, with no edge states. However, for $\mu<1$ the system has a topological massive Dirac fermion at the edges, as shown in the wave function plot in Fig.~\ref{Top_Dirac}(b). The two Majorana modes at the two distant edges have paired up onto a single massive Dirac fermion. Notice that the fermion is highly non-local and its nature is deeply rooted in the long-range/non-local character of the pairing term (see Sec. III of the Appendix for details). We notice that if we had considered imaginary pairing amplitudes within $D$ symmetry class (particle-hole symmetric), the non-local massive Dirac fermions would persist. This topological quasi-particle is still protected by fermion parity: the ground state has still even parity, whereas the first excited state populates this non-local massive fermion and has odd parity. One cannot induce a transition between these two states without violating the fermion parity conservation of the Hamiltonian, and applying a non-local operation is needed. Moreover, the subspace of these two edge states is still protected by the bulk gap from bulk excitations. The conservation of fermion parity and the non-local character of the massive Dirac fermion make these two states ideal to define a topological qubit using two copies of the Kitaev chain \cite{rnpj1,Bravyi_Kitaev02,Alicea_et_al_11,Baranov_et_al_13}. Further details are detoured to Sec.V of the SM. Additionally, in Sec.~II of the Appendix, by means of finite-size scaling we show that the mass of the Dirac fermion stays finite in the thermodynamic limit for $\mu<1$ and $\alpha<1$. This way we can prove that the effect is purely topological and caused by the long-range deformation. 

When we close the chain, the edge states disappear as we may expect for a topological effect. Despite the long-range pairing coupling, the system still belongs to the BDI symmetry class \cite{Ludwig,Kitaev_2009}, since no discrete symmetry has been broken.
The winding number $\omega$ can still be formally defined using Eq.~\eqref{w}. However, the topological singularity at $k=0$ deeply modifies the value of $\omega$. For the trivial phase $\mu>1$, the winding number is $\omega=-1/2$, whereas for the new unconventional topological phase is $\omega=+1/2$ if $\mu<1$. The semi-integer character of $\omega$ is associated to the integrable divergence at $k=0$, which modifies the continuous mapping $\mathit{S}^1\longrightarrow \mathit{S}^1$. 
Notwithstanding, in this region there is still a jump of one unit between the two topologically different phases,
 $\Delta \omega=\omega_{{\rm top}}-\omega_{{\rm trivial}}=1$  (see Fig.~\ref{pairing_phase}). Moreover, the topological indicators take on the same value within the whole phase until the bulk gap closes at $\mu=1$, giving rise to a topological phase transition, and the new massive topological edge states disappear. Therefore, we can still establish a bulk-edge correspondence. 

There is a novelty in this case regarding the parallel transport for the Berry phase. Namely, at $k=0$ the adiabatic condition breaks down since both the energy dispersion relation $E_k$ and the quasi-particle group velocity $\partial_kE_k$ diverge. Moreover, the singularity at $k=0$ of the lower band eigenvector $\ket{u^{-}_k}$, cannot be removed by a simple gauge transformation as it is not just a $U(1)$ phase difference, but a phase shift unitary jump,
\begin{equation}
\ket{u^{-}_{k\rightarrow0^+}}={\rm e}^{{\rm i}\pi P_{\pm}}\ket{u^{-}_{k\rightarrow0^-}},
\label{utrans2}
\end{equation} 
where $P_{\pm}=\frac{1}{2}\big(\mathds{1}\pm\sigma_z\big)$. More explicitly,
\begin{equation}
{\rm e}^{{\rm i}\pi P_{-}}=\begin{pmatrix} 1 & 0 \\ 0 & {\rm e}^{{\rm i}\pi} \end{pmatrix},~~{\rm e}^{{\rm i}\pi P_{+}}=\begin{pmatrix} {\rm e}^{{\rm i}\pi} & 0 \\ 0 & 1 \end{pmatrix}.
\end{equation}
The difference in sign $\pm$ of the projector $P_{\pm}$ depends on the topological regime. For $\mu>1$, the system is in a trivial phase with no edge states and the long-range singularity of $\ket{u^{-}_{k}}$ at $k=0$ is given by ${\rm e}^{{\rm i}\pi P_{-}}$. On the other hand for $\mu<1$, the system is in a topological phase with massive and non-local edge states. The singularity of $\ket{u^{-}_{k}}$ at $k=0$ in that case is given by ${\rm e}^{{\rm i}\pi P_{+}}$. 

iii/ Crossover Sector [$\alpha\in(1,{3}/{2})$]--- this is a crossover region between sectors i/ and ii/. Within this sector, there are massless Majorana edge states for $-1<\mu<1$ like in sector i/, but for $\mu<-1$ the edge states become massive like in sector ii/. This is shown through finite-size scaling in Sec.~II and III of the Appendix. The intuition behind this result is that the gap closes in the thermodynamic limit at $\mu=-1$ also for $\alpha\in(1,\frac{3}{2})$. The dispersion relation $E_k$ is no longer divergent, however its derivative $\partial_kE_k$ (the group velocity) is still singular at $k=0$ and the structure of the topological singularity changes accordingly.
The winding number is not able to capture the ``mixed'' character of this sector. However, as detailed in Sec.~I of the Appendix, we can clearly see that the behaviour of the winding vector and the lower-band eigenstate is different from the other two sectors.

In Fig.~\ref{pairing_phase}, we present a complete phase diagram summarising the different topological phases of the model as a function of $\mu$ and $\alpha$.

\noindent {\it 5. Outlook and Conclusions.}
We have found that finite-range and long-range extensions of the one-dimensional Kitaev chain can be used as a resource for enhancing existing topological properties and for unveiling new topological effects. In particular, for long-range pairing deformations, we observe non-local massive Dirac fermions characterised by fractional topological numbers. Hamiltonians with long-range pairing and hopping may be realised in Shiba chains as recently proposed in~\cite{Pientka2013,Pientka2014}, where edge states can be detected, e.g., by scanning tunneling spectroscopy~\cite{Yazdan1997}. Alternatively, next-nearest neighbour hopping may be harnessed in atomic and molecular setups~\cite{Jian_et_al11}, where massive edge modes should be observable via a combination of spectroscopic techniques and single-site addressing~\cite{Bakr2009,Sherson2010}. The extension of existing models for qubits, constructed by
topologically protected gapped modes, may boost the search for long-range deformations in more complicated topological models with symmetry-protected or even intrinsic topological order.

\begin{acknowledgments}

M.A.MD. and O.V. thank the Spanish MINECO grant FIS2012-33152, the CAM research consortium QUITEMAD+ S2013/ICE-2801, the U.S. Army Research Office through grant W911NF-14-1-0103, FPU MECD Grant and Residencia de Estudiantes. G.P. and D.V. acknowledge support by the ERC-St Grant ColdSIM (No. 307688),  EOARD,  UdS  via  Labex  NIE, ANR via BLUSHIELD  and  IdEX,  RYSQ.

\end{acknowledgments}

\clearpage

\section{APPENDIX}

\appendix

\setcounter{figure}{0}
\setcounter{equation}{0}
\renewcommand*{\thefigure}{A\arabic{figure}}
\renewcommand*{\theequation}{A\arabic{equation}}

\subsection{I.\quad Winding vector and Berry phase in the presence of a Topological Singularity}
\label{app_A}

\begin{figure}[t]
\centering
\includegraphics[width=\columnwidth]{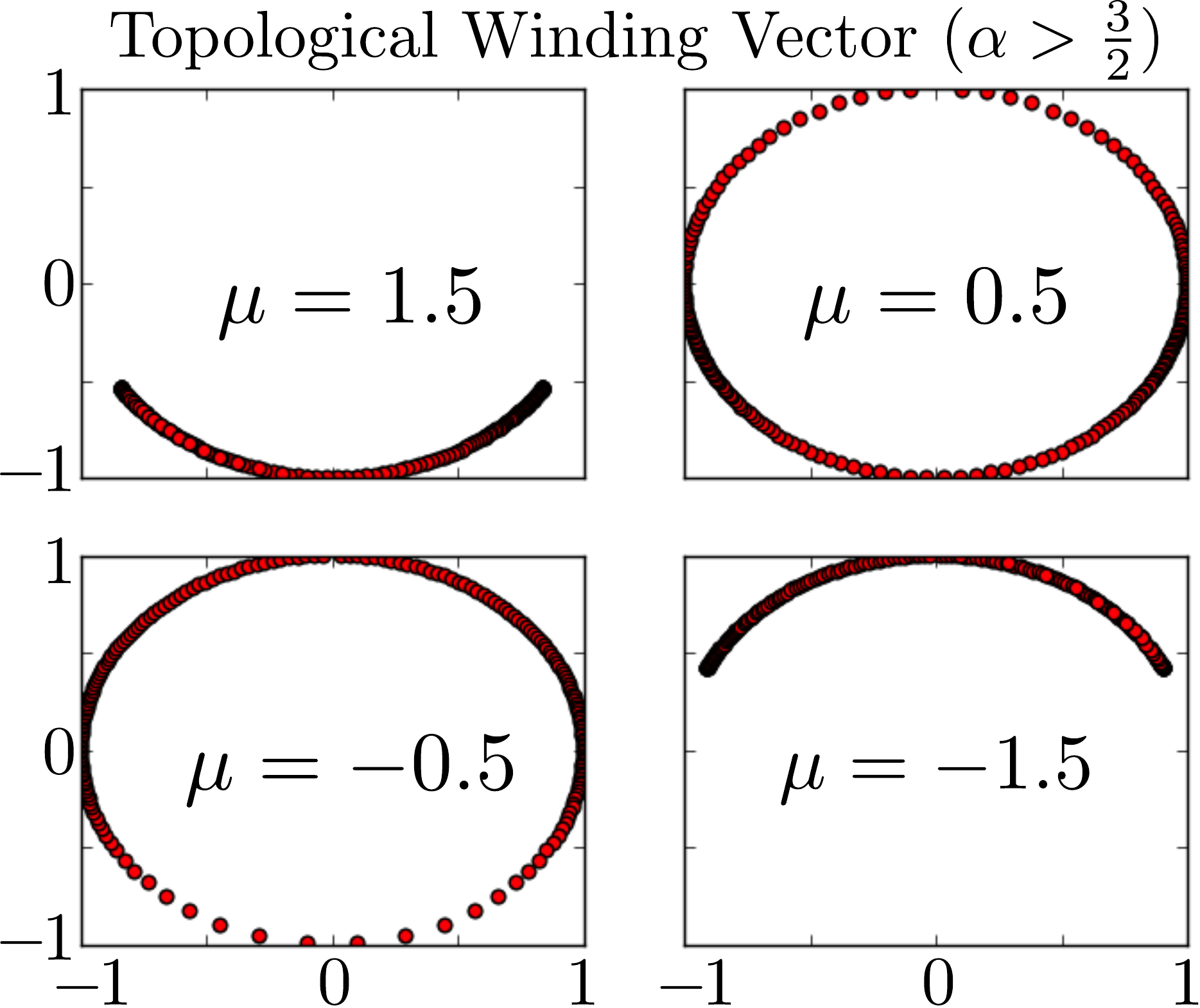}
\caption{Trajectories of the winding vector for different regions within the Majorana sector, for $L=201$ sites and $\alpha=3$. The red spots represent the movement of the winding vector along the unit circle $S^1$. As we see, for $\mu>1$ and $\mu<-1$, the vector never winds around the whole $S^1$, just moving back and forth twice. However, if $\mu\in(-1,1)$ the vector winds around $S^1$. The darker regions highlight a larger density of points.}
\label{SM_winding_p_a3p0}
\end{figure}

In the thermodynamic limit $L\rightarrow\infty$, the function $f_{\alpha}(k)$ in (3) 
tends to
\begin{equation}
f_{\alpha}(k)=\frac{{\rm i}}{2}\Big({\rm Li}_{\alpha}({\rm e}^{-{\rm i}k})-{\rm Li}_{\alpha}({\rm e}^{{\rm i}k})\Big).
\label{fk}
\end{equation}
where ${\rm Li}_{\alpha}({\rm e}^{{\rm i}k})$ is a polylogarithmic function. This function defines the long-range pairing and is divergent at $k=0$ for $\alpha<1$. As a consequence, the energy gap in (2) 
diverges if $\alpha<1$, and the particle group velocity $v_g=\partial_kE_k$ diverges for $\alpha<{3}/{2}$ at $k=0$. In line with this,  we can trace the effect of this divergence over the winding vector and the topological phases within the different topological sectors:
\vspace{0.1cm}

i/ Majorana Sector [$\alpha>{3}/{2}$]--- There is no topological singularity in this sector. The winding number and the Berry phase can be computed using (4) 
and (5). 
We find the same topological indicators and the same type of edge states physics as for the short range Kitaev model ($\alpha\rightarrow\infty$). In Fig.~\ref{SM_winding_p_a3p0} we plot the winding vector for $L=201$ sites and for different values of the chemical potential $\mu$ belonging to different topological regimes. We can see that for trivial regions $|\mu|>1$, the winding vector winds back and forth and never covers the entire $S^1$ circle. On the other hand, when the system is within a topological phase $\mu\in(-1,1)$, the winding vector winds around $S^1$ completely. 
The lower band eigenvector $\ket{u^{-}_k}$ can always be chosen to be periodic. If the system is in the trivial phase, $\ket{u^{-}_k}$ is also continuous, whereas inside the topological phase, there is a $U(1)$ phase discontinuity at $k=0$, i.e.,
\begin{equation}
\ket{u^{-}_{k\rightarrow0^+}}={\rm e}^{{\rm i}\pi} \ket{u^{-}_{k\rightarrow0^-}}.
\label{ut1}
\end{equation} 
This phase shift can be gauged away from $k=0$, and it represents the Berry phase gained by the system after an adiabatic transport from a certain crystalline momentum $k_0$ up to $k_0+G$, where $G$ is a reciprocal lattice vector.

\begin{figure}[t]
\centering
\includegraphics[width=\columnwidth]{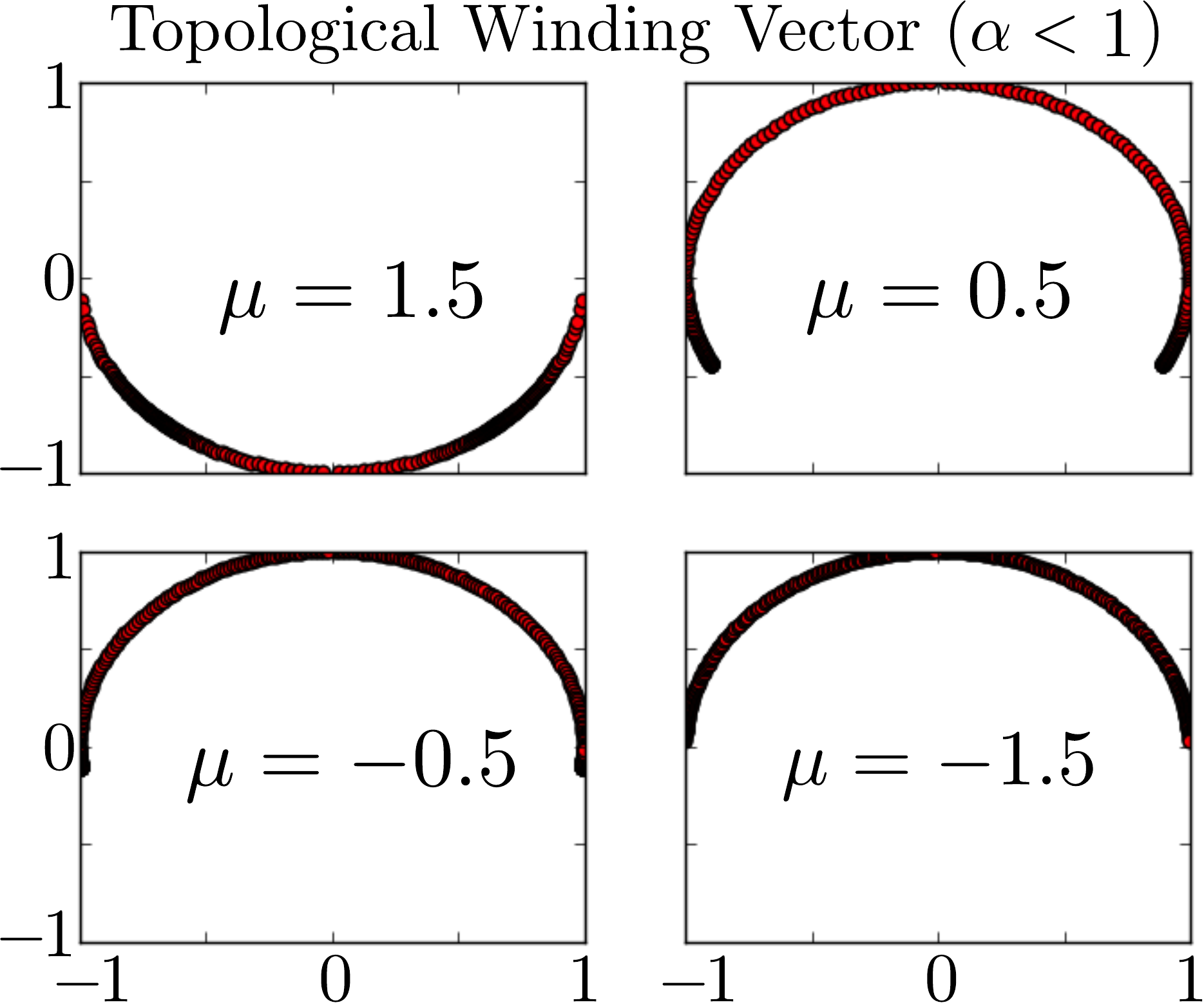}
\caption{Trajectories of the winding vector for different regions within the massive Dirac sector, for $L=301$ sites and $\alpha=0.5$. The red spots represent the movement of the winding vector along the unit circle $S^1$. For $\mu>1$, the winding vector covers the lower half of $S^1$. However, if $\mu<1$ the vector covers only the upper half of the circle. The darker regions highlight a larger density of points.}
\label{SM_winding_p_a0p5}
\end{figure}

\vspace{0.2cm}

ii/ Massive Dirac Sector [$\alpha<1$]--- The topological singularity at $k=0$ makes the winding vector ill-defined at that point, although its contribution to the winding number can still be integrated. In Fig.~\ref{SM_winding_p_a0p5} we plot the winding vector for $L=301$ sites and for different values of $\mu$. In particular, for $\mu>1$ the winding vector covers the entire lower half of the $S^1$ circle, explaining the value $\omega=-{1}/{2}$ of the winding number. On the contrary, for $\mu<1$, the winding vector just covers the entire upper half of $S^1$ as shown in the figure. The function $f_{\alpha}(k)$ at $k=0$  diverges as
\begin{equation}
f_{\alpha}(k\rightarrow0^-)\longrightarrow-\infty~,~f_{\alpha}(k\rightarrow0^+)\longrightarrow\infty.
\label{alpha_limits}
\end{equation}
Hence, in the transition from $k<0$ to $k>0$, the winding vector skips the entire lower part of the $S^1$ circle because of the topological singularity at $k=0$. This explains the value of the winding vector $\omega=+{1}/{2}$ in this new topological phase.
Complementary, the adiabatic condition breaks down at $k=0$ as the quasi-particle group velocity diverges. Therefore, we can no longer say that the system picks up a $U(1)$ phase after a close loop in momentum space. Actually, the singularity at $k=0$ of the lower band eigenvector $\ket{u^{-}_k}$, cannot be removed by a simple gauge transformation as it is not just a $U(1)$ phase difference, but a phase shift unitary jump,
\begin{equation}
\ket{u^{-}_{k\rightarrow0^+}}={\rm e}^{{\rm i}\pi P_{\pm}}\ket{u^{-}_{k\rightarrow0^-}},
\label{utrans2}
\end{equation} 
where $P_{\pm}=\frac{1}{2}\big(\mathds{1}\pm\sigma_z\big)$. More explicitly,
\begin{equation}
{\rm e}^{{\rm i}\pi P_{-}}=\begin{pmatrix} 1 & 0 \\ 0 & {\rm e}^{{\rm i}\pi} \end{pmatrix},~~{\rm e}^{{\rm i}\pi P_{+}}=\begin{pmatrix} {\rm e}^{{\rm i}\pi} & 0 \\ 0 & 1 \end{pmatrix}.
\end{equation}
The difference in sign $\pm$ of the projector $P_{\pm}$ depends on the topological sector. For $\mu>1$, the system is in a trivial phase with no edge states and the singularity of the eigenstate at $k=0$ is given by ${\rm e}^{{\rm i}\pi P_{-}}$. On the other hand for $\mu<1$, the system is in a topological phase with massive and non-local edge states. The singularity of the eigenstate at $k=0$ in that case is given by ${\rm e}^{{\rm i}\pi P_{+}}$.

\vspace{0.2cm}

iii/ Crossover Sector [$\alpha\in(1,{3}/{2})$]--- This is a crossover region between the previous two sectors i/ and ii/. In this case, the structure of the topological singularity at $k=0$ has changed. The components of the winding vector are continuous, but their derivatives diverge. Therefore, the population of points close to $k=0$ is extremely dispersive.  The winding vector does not cover the entire south pole sector due to the divergence in the derivatives of its components in the thermodynamic limit. Hence, the behaviour of the winding vector is different than in the other two previous sectors i/ and ii/. This might be linked to the mixed character of the sector with the presence of MZMs ($-1<\mu<1$) and massive Dirac fermions ($\mu<-1$). Regarding the lower band eigenvector, it is continuous but its derivative is still divergent at $k=0$ breaking the adiabatic condition.

\subsection{II.\quad Edge mass gap finite-size scaling}
\label{app_B}

In Sec.~IV of the main text, we claimed that the pairing of the MZMs into a massive non-local Dirac fermion cannot be explained as a simple interaction between the Majorana fermions at the edges due to a finite size effect. Actually, its nature is deeply rooted into the long-range/non-local character of the pairing deformation of the Kitaev chain. The absence of a degenerate zero energy subspace avoids a wave function superposition to localise a single Majorana mode at one edge only. On the contrary, the two edges are inevitably coupled to each other, pairing to a non-local massive Dirac mode as shown in Fig. 2.

In order to proof this claim more rigorously, we have computed the mass of the edge states through a finite-size scaling for different values of the decaying exponent $\alpha$ and the chemical potential $\mu$.

\begin{figure}[t]
\centering
\includegraphics[width=\columnwidth]{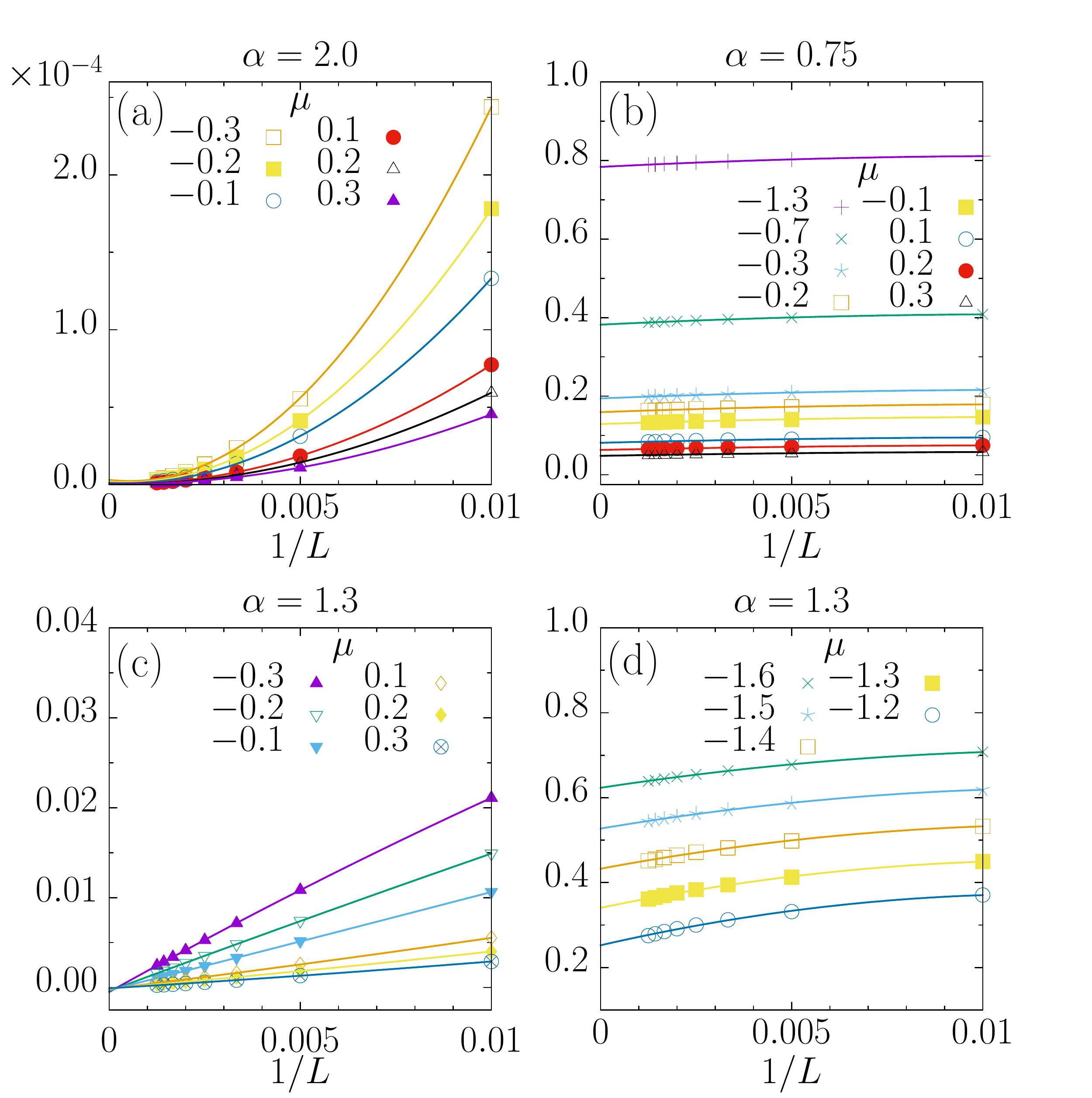}
\caption{Finite size scaling of the edge mass gap for different values of the exponent $\alpha$ and the chemical potential $\mu$. In~(a) we plot the case for $\alpha=2.0$ belonging to the Majorana sector. The edge mass gap closes with $L$ for every $\mu$ within the topological phase. In~(b) we take $\alpha=0.75$ in the massive Dirac sector. For $\mu<1$ the edge mass gap tends to a finite value in the thermodynamic limit. In~(c) and (d) we perform the finite-size scaling for $\alpha=1.3$ within the crossover sector. If $-1<\mu<1$ (c), there are massless edge states. If $\mu<-1$ (d), there are massive edge states up to our numerical precision.}
\label{SM_scaling_all}
\end{figure}

In Fig.~\ref{SM_scaling_all}(a) we perform a finite-size scaling for the masses of the MZMs for the Majorana sector. Within the topological sector $\mu\in(-1,1)$, the edge mass gap clearly goes to zero with $L$ as we expected. In Fig.~\ref{SM_scaling_all}(b) we perform the same finite-size scaling analysis for the massive Dirac sector. In this case, there are edge states for $\mu<1$. As we can see, the masses of the edge states depend on both $\mu$ and $\alpha$, and go to a finite value even in the thermodynamic limit. This proves that the topological nature of the non-local massive Dirac fermions purely comes from the long-range deformation of the original Kitaev Hamiltonian and not from a finite size effect. 

On the other hand, Fig.~\ref{SM_scaling_all}(c) and Fig.~\ref{SM_scaling_all}(d) show the finite-size scaling for the edge mass gap within the crossover sector. Although there are edge states all over $\mu<1$, they can be either massive or massless depending on the chemical potential $\mu$. If $-1<\mu<1$ the edge states are massless as shown in Fig.~\ref{SM_scaling_all}(c), whereas for $\mu<-1$ the edge states become massive as shown in Fig.~\ref{SM_scaling_all}(d). Hence, this sector displays a mixed character between a Majorana and a massive Dirac phase.

\begin{figure}[t]
\begin{center}
\includegraphics[width=0.4\textwidth]{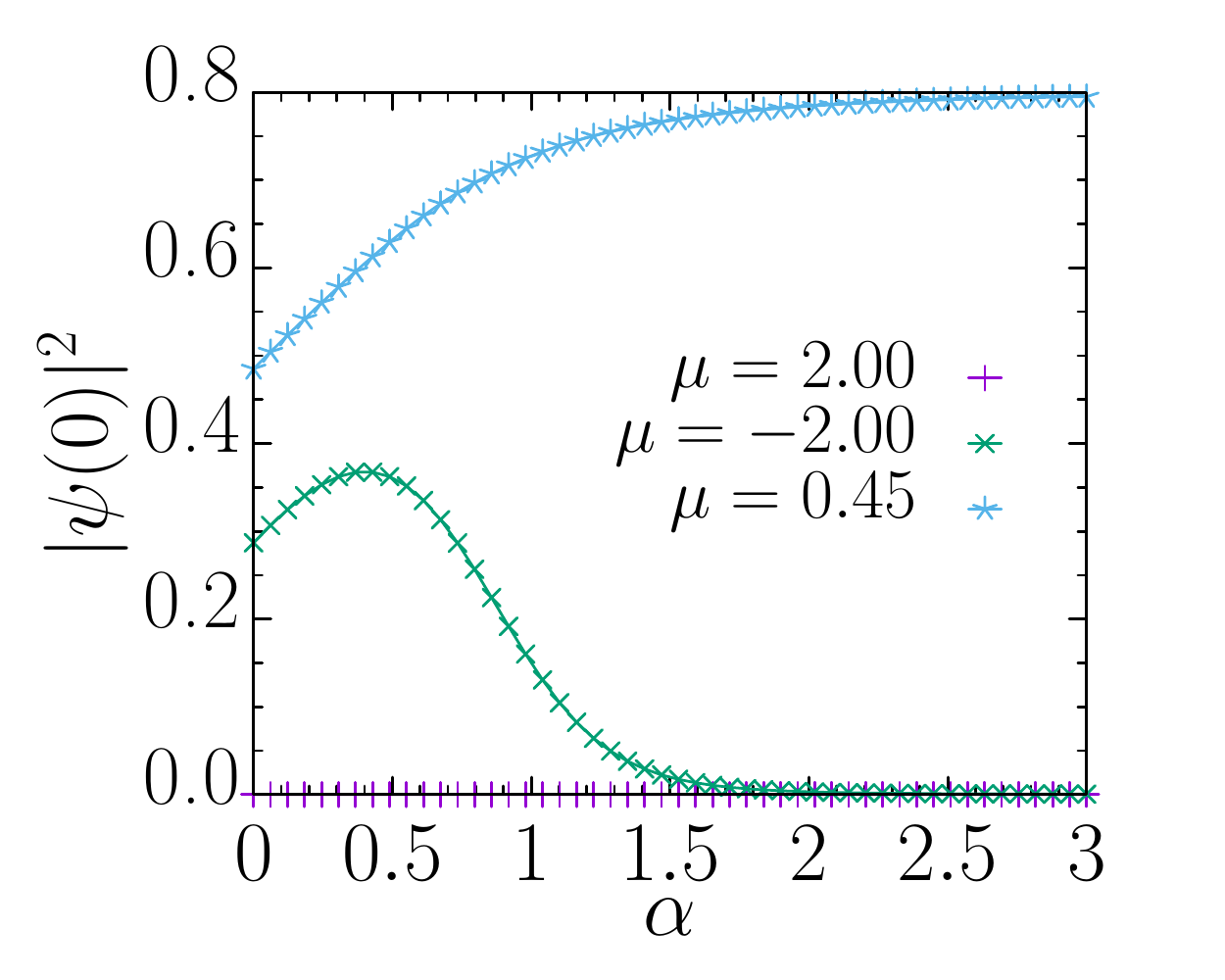} 
\end{center}
\caption{We plot the wavefunction probability density $|\psi(0)|^2$ for the lowest-energy eigenstate at one of the edges for different $\mu$ as a function of $\alpha$: i/ $\mu>1$ (purple line) there are no edge states regardless of $\alpha$. ii/ $-1<\mu<1$ (blue line) there is always a finite edge-state density. iii/ $\mu<-1$ (green line) there are edge states if $\alpha\lesssim\frac{3}{2}$.} 
\label{edge_density}
\end{figure}

Lastly, we have investigated the edge properties of the model by studying the wavefunction probability density $|\psi(0)|^2$ for the lowest-energy single-particle eigenstate at one of the edges.
Fig.~\ref{edge_density} shows $|\psi(0)|^2$  for different $\mu$ as a function of $\alpha$. The results can be summarised as follows: i/ for $\mu>1$ (purple line) there are no edge states regardless of $\alpha$. ii/ If $-1<\mu<1$ (blue line) there is always a finite edge-state density. In addition, from Fig.~\ref{SM_scaling_all} we obtain that if $\alpha>1$ the edge states are massless, whereas if $\alpha<1$ they are massive. iii/ If $\mu<-1$ (green line) there are edge states if $\alpha<\frac{3}{2}$. Then, from Fig.~\ref{SM_scaling_all} we conclude that they are always massive in this case. Actually, we can even monitor how one of the Dirac bulk states gets transmuted into a non-local massive Dirac edge mode by lowering $\alpha$.

The results are in complete agreement with Fig.3 from the main text and the finite size scaling analysis in Fig.~\ref{SM_scaling_all}.

\subsection{III.\quad Analytical Structure of the Edge States}
\label{app_C}

We take the Hamiltonian defined in Eq. (1) of the main text, where only long-range pairing terms are considered. For the parameter choice $\theta=0$ and $J=|M|=\frac{1}{2}$ assumed along the paper, we can separate the short-range and long-range contributions:
\begin{equation}
H:=H_{SR}+H_{\mu}+H_{LR}, 
\end{equation}
where
\begin{eqnarray}
\label{H2}
H_{SR}&:=&\sum^L_{j=1}\left(-a^{\dagger}_ja_{j+1}+a_ja_{j+1}+\text{h.c}\right),\nonumber\\ 
H_{\mu}&:=&-\mu\sum^L_{j=1}\left(a_j^{\dagger}a_j-\frac{1}{2}\right),\nonumber\\
H_{LR}&:=& \sum^L_{j=1}\sum_{l=2}^{L-j}\frac{1}{R_{l,\alpha}}a_ja_{j+l}+\text{h.c}
\end{eqnarray}

In order to uncover the different topological phases of the model, we will rewrite this Hamiltonian in terms of Majorana operators  
\begin{equation}
\label{majorana_op}
c_{j}=\textstyle{\frac{1}{\sqrt{2}}}\left(a^{\dagger}_j+a_j\right),~~~~~d_{j}=\textstyle{\frac{\ii}{\sqrt{2}}}\left(a^{\dagger}_j-a_j\right), 
\end{equation}
satisfying the fermionic anticommutation relation $\{c_{i},c_j\}=\delta_{ij}$, but also the Majorana condition $c^{\dagger}_j=c_j$ and the same for $d_j$. 

Substituting Eq.\eqref{majorana_op} in Eq.\eqref{H2}, we get
\begin{eqnarray}
\label{H3}
H_{SR}&=&{\rm i}\sum^{L-1}_{j=1}d_{j}c_{j+1},\nonumber\\ 
H_{\mu}&=&-{\rm i}\mu\sum^{L}_{j=1}c_{j}d_{j},\nonumber\\
H_{LR}&=& \frac{{\rm i}}{2} \sum_{j=1}^L\sum_{l=2}^{L-j}\frac{1}{R_{l,\alpha}}\Big(d_jc_{j+l}+c_jd_{j+l}\Big). 
\end{eqnarray}

\subsection{Short-range Kitaev chain}

Let us first consider the purely short-range Kitaev chain with a chemical potential. Majorana fermions usually pair locally in such a way that they constitute a regular Dirac fermion. However, for open boundary conditions and certain values of the coupling constants, two Majorana fermions at the boundary remain unpaired. The Majorana operators at the edges $c_1$ and $d_{N}$ do not appear in the short-range Hamiltonian $H_{SR}$, hence, if we set $\mu=0$ they become zero energy modes (see Fig. 1(b) in the main text) as they decouple from the dynamics. These MZMs are topologically protected and represent a hallmark of topological order in the system. 

Note that if $\mu\not=0$, then $c_1$ and $d_{N}$ do appear in $H_{\rm SR}+H_{\mu}$, however we will show that these modes have an exponentially small energy and are exponentially localised at the edges. Hence, in the thermodynamic limit they become exact zero energy modes.

In order to prove this, we will elaborate on an ansatz method \cite{Fendly_12} in order to construct the edge modes analytically for the short-range Kitaev chain. A fermionic zero mode $\Psi$ is an operator that commutes with the Hamiltonian: $[H,\Psi]=0$, anticommutes with $(-1)^F$, i.e.$~~\{(-1)^F,H\}=0$ and it is conveniently normalised.
The second property guarantees that the operator $\Psi$ maps the odd and even parity sectors. The first one instead imposes the condition for a zero energy mode, based on the Heisenberg equation $\frac{d\Psi}{dt}=-{\rm i}[H,\Psi]$.\\

First of all, we note that the fermionic mode constructed out of the two unpaired Majoranas $c_1$ and $d_N$, namely, 
\begin{equation}
\tilde{a}_{\rm E}=\frac{1}{\sqrt{2}}(c_1+{\rm i}d_{N}),
\label{ae}
\end{equation}
clearly commutes with $H_{\rm SR}+H_{\mu}$ for $\mu=0$, $[H_{\rm SR}+H_{\mu},\tilde{a}_{\rm E}]=0$. Moreover, $\tilde{a}_{\rm E}$ destroys a fermionic mode in the system, thereby mapping the even and odd parity sectors. For $\mu\not=0$ instead, the fermionic operator $\tilde{a}_{\rm E}$ does not conmute with $H$, however, the new edge mode can still be determined.

Let us propose an ansatz wavefunction for the modified left Majorana mode $\Phi_{\rm left}$. We know that $\Phi_{\rm left}(\mu=0)=c_1$ and the Hamiltonians defined in Eq. \eqref{H3}, $H_{\rm SR}$, $H_{\mu}$, $H_{LR}$ have only mixing terms $c_{i}d_{j}$. Hence, the most general ansatz would be:
\begin{equation}
\Phi_{\rm left}=\sum_{j=1}^Lm_jc_j
\label{phil1}
\end{equation}
where $c_j$ are Majorana operators and $m_j$ are real coefficients to be determined. Namely, 

\begin{equation}
[H_{SR}+H_{\mu},\Phi_{\rm left}] = {\rm i}\sum_{j=1}^{L-1}\Big(m_{j+1}+\mu m_j\Big)d_j~+\mu m_Ld_L,
\label{ComSR}
\end{equation}
where we have used the anticonmuting properties of the Majorana operators, 
\begin{equation}
[d_kc_j,c_l]=\delta_{j,l}d_k,~~~\text{and}~~ [c_jd_k,c_l]=-\delta_{j,l}d_k.
\end{equation}

\begin{figure}[t]
\centering
\includegraphics[width=\columnwidth]{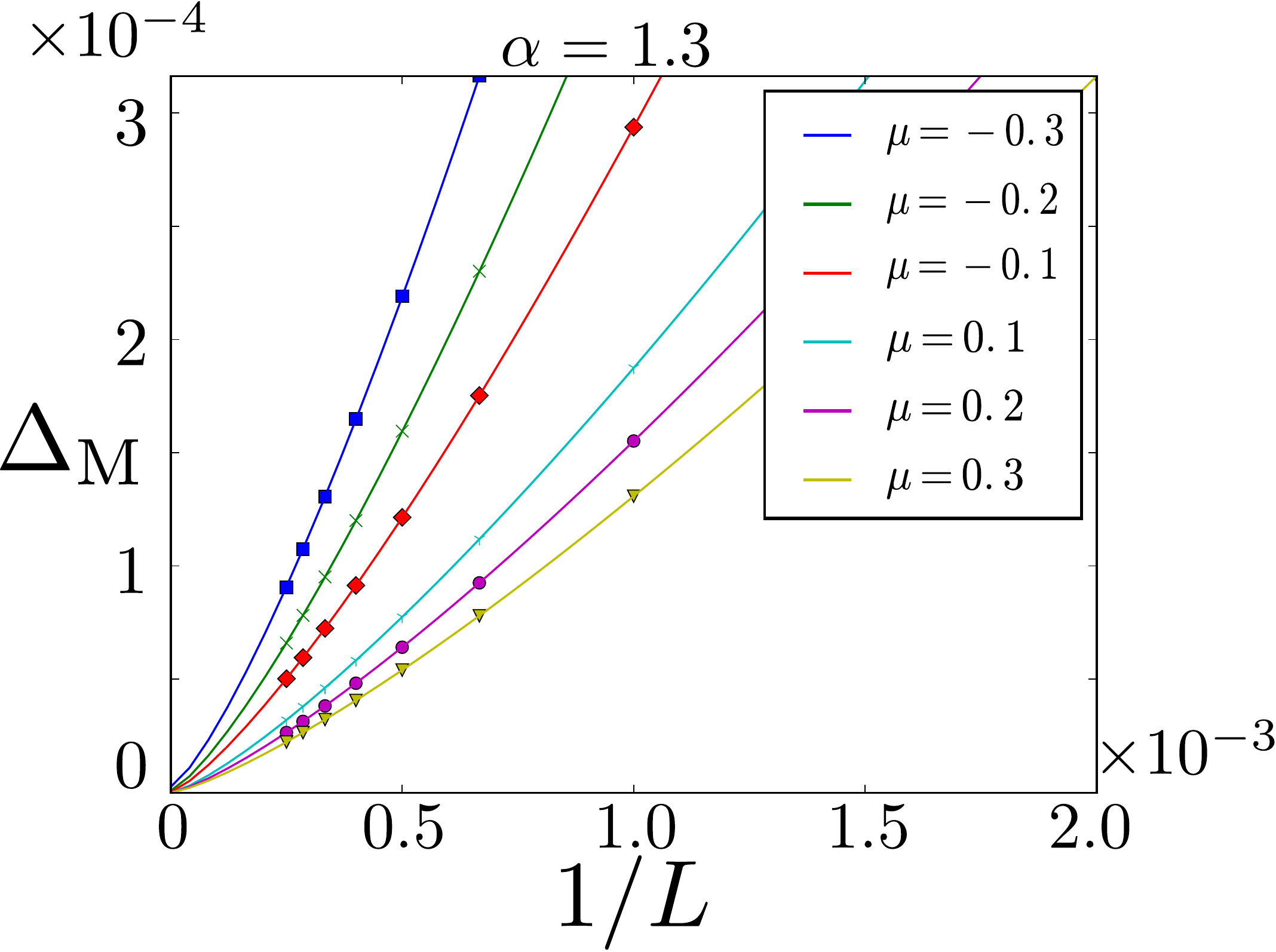}
\caption{Finite size scaling of the mass indicator $\Delta_{M}$ within the crossover sector, $\alpha\in(\frac{3}{2},1)$, for the massless phase ($-1<\mu<1$) where we expect MZMs. As we can see, $\Delta_{M}$ goes to zero when $L$ increases.} 
\label{SM_scaling_Delta}
\end{figure}

If we want to make $\Phi_{\rm left}$ a MZM, then we should impose the commutator in Eq. \eqref{ComSR} to be zero. Hence,
\begin{equation}
m_{j+1}+\mu m_j=0~~~~\forall j=1,...,L-1.
\label{rec1}
\end{equation}
Note that the coefficient accompanying $d_L$ will be determined with the $L-1$ previous equations. 
For continuity with the $\mu=0$, we take $m_1=1$ up to normalisation of the final wavefunction. It is now very easy to see that the solution to the recursive equation \eqref{rec1} is
\begin{equation}
m_j=(-\mu)^{j-1}~~~\forall j=2,...,L.
\end{equation}
Thus,
\begin{equation}
\Phi_{\rm left}=c_1- \mu c_2 + \mu^2 c_3 - \mu^3c_4 + ...
\end{equation}
and
\begin{equation}
[H_{SR}+H_{\mu},\Phi_{\rm left}] = \mu m_Ld_L=\mu(-\mu)^{L-1}d_L.
\end{equation}
The same equation holds for the right edge Majorana mode,
\begin{equation}
\Phi_{\rm right}=d_{L} - \mu d_{L-1} + \mu^2 d_{L-2} - \mu^3d_{L-3} + ...
\end{equation}
Hence, the two new Majorana (almost) zero modes localised around the left and right edges for $|\mu|<1$, can be combined into a Dirac fermionic edge mode, $\Psi_{\rm E}=\frac{1}{\sqrt{2}}\Big(\Phi_{\rm left}+{\rm i}\Phi_{\rm right}\Big)$. This edge fermion doesn't commute exactly with $H_{\rm SR}+H_{\mu}$,
\begin{equation}
[H_{SR}+H_{\mu},\Psi_{\rm E}]=\mu(-\mu)^{L-1}\frac{1}{\sqrt{2}}(c_{1}+{\rm i} d_{L}).
\end{equation}
However, this coefficient is exponentially small in $L$ as long as $|\mu|<1$. Hence, in the thermodynamic limit $L\rightarrow\infty$, the fermionic mode $\Psi_{\rm E}$ commutes with the Hamiltonian, satisfying the condition to be a zero energy mode, and the new Majorana fermionic operators $\Phi_{\rm left}$ and $\Phi_{\rm right}$ are unpaired.\\

\begin{figure}
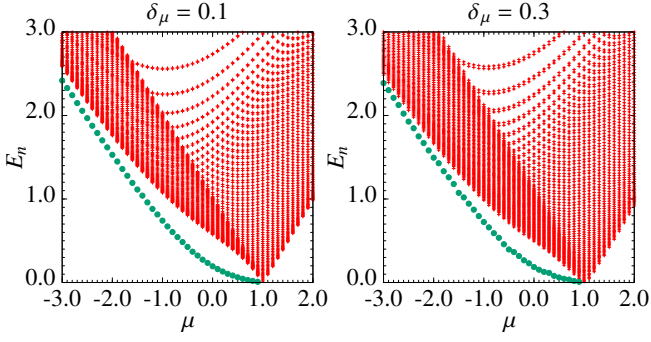

\includegraphics[scale=0.35]{{{energies_eps_0.1}}}%
\includegraphics[scale=0.35]{{{energies_eps_0.3}}}%
\caption{The plot shows the energy spectrum of the  Hamiltonian with long-rang pairing Eq.~(1) of the main text and a random chemical potential term Eq.~\eqref{ham_disorder} as a function of the chemical potential $\mu$ for two values of the disorder strength~$\delta_\mu$. The states plotted in green are separated by a finite gap from the band of bulk states plotted in red.}\label{fig_disorder}
\end{figure}

\subsection{Kitaev chain with long-range couplings.}

Let us now include the long-range deformations given by $H_{LR}$, 
\begin{equation}
\begin{split}
&[H_{LR},\Phi_{\rm left}]  = \\
&=\frac{{\rm i}}{2}\sum_{j=1}^{L}\sum_{l=2}^{L-j}\sum_{k=1}^{L}\frac{m_k}{R_{l,\alpha}}\Big( [d_jc_{j+l},c_k]
+[c_jd_{j+l},c_k]\Big)  = \\
&=\frac{{\rm i}}{2}\Big(\sum_{j=1}^{L-2}\sum_{l=2}^{L-j}\frac{m_{j+l}}{R_{l,\alpha}}d_j-\sum_{j=1}^{L-2}\sum_{l=2}^{L-j}\frac{m_{j}}{R_{l,\alpha}}d_{j+l}\Big).
\label{HLRcom}
\end{split}
\end{equation}
The commutator of total Hamiltonian, $H=H_{SR}+H_{\mu}+H_{LR}$, can be regrouped into a similar fashion as for the short-range case, but with more complicated contributions,
\begin{eqnarray}
&&[H,\Phi_{\rm left}] = {\rm i}\sum_{j=1}^{L-1}\Big(m_{j+1}+\mu m_j +\sum_{l=2}^{L-j}\frac{m_{j+l}}{2R_{l,\alpha}} -\nonumber\\
&-&\sum_{k=1}^{j-2} \frac{m_k}{2R_{j-k,\alpha}}\Big)d_j+{\rm i}\Big(\mu m_L - \sum_{k=1}^{L-2} \frac{m_k}{2R_{L-k,\alpha}} \Big)d_L.\nonumber\\
\label{Hcom}
\end{eqnarray}

Proceeding as in the previous section, we now impose all the coefficients accompanying the operators $d_j$ to be zero, except the one coming from $d_L$, which will be automatically determined by these equations. 
Then, if the coefficient accompanying $d_L$ goes to zero as $L$ increases, we can claim that we still have unpaired Majorana modes. Otherwise, a massive non-local Dirac mode will appear. 

Therefore, we need to solve the following discrete equation:
\begin{equation}
m_{j+1}+\mu m_j +\sum_{l=2}^{L-j}\frac{m_{j+l}}{2R_{l,\alpha}} - \sum_{k=1}^{j-2} \frac{m_k}{2R_{j-k,\alpha}}=0,~~~\forall j=1,...,L-1
\label{rec2}
\end{equation}
Note that the first sum only contributes from $j=3$ and the last sum only runs up to $j=L-2$.
Once we solve Eq. \eqref{rec2}, the edge mass indicator of the system is determined from Eq. \eqref{Hcom}, by computing
\begin{equation}
\Delta_{M}:=\mu m_L - \sum_{k=1}^{L-2} \frac{m_k}{2R_{L-k,\alpha}}
\label{De}
\end{equation}
Note that $\Delta_{M}$ is not exactly the edge mass gap, however, it distinguishes the region where we have MZM or when they turn into a non-local massive Dirac fermion, $[H,\Psi_{\rm E}]=\Delta_{M}\frac{1}{\sqrt{2}}(c_1+{\rm i}d_L)$. If $\Delta_{M}$ goes to zero as $L$ increases, then we will certainly have a MZM. Note the analogy with the short-range case with a non-zero chemical potential $\mu$.

Although a complete analytical solution might be involved, this process can be easily programmed in a computer as a set of linear equations. In Fig. \ref{SM_scaling_Delta}, we compute the finite-size scaling of $\Delta_{M}$ within the crossover sector for $-1<\mu<1$ where we expect MZMs. Up to our numerical precision, $\Delta_{M}$ goes to zero in perfect accordance with the phase diagram of Fig. 3 in the main text and the finite-size scaling for the energy of the edge modes in Fig. \ref{SM_scaling_all}. 

Before concluding, we would like to give an intuitive picture to explain the mechanism that pairs MZMs non-locally via the long-range coupling. At first sight, one might think that long-range interactions would couple every Majorana fermion with each other, mixing them all. However, $c_1$ and $d_N$ commute with $H_{SR}={\rm i}\sum^{L-1}_{j=1}d_{j}c_{j+1}$, and the long-range Hamiltonian $H_{LR}$ only couples the two of them together (up to exponential and algebraic tails). This can be indeed inferred from the commutator of $\tilde{a}_{\rm E}=\frac{1}{\sqrt{2}}(c_1+{\rm i}d_N)$ and $H_{LR}$:
\begin{equation}
[H_{LR},\tilde{a}_{\rm E}]=-\frac{1}{2R_{L-1,\alpha}}\tilde{a}_{\rm E} - \sum_{j=2}^{L-2} \frac{1}{2R_{L-j,\alpha}}\tilde{a}_j,
\label{aEc}
\end{equation}
where $\tilde{a}_j=\frac{1}{\sqrt{2}}(c_j+{\rm i}d_{L+1-j})$ are new bulk fermionic modes, and the edge Majoranas $c_1$ and $d_N$ only appear in $\tilde{a}_E$.

Actually, a complementary way to construct the new fermionic edge mode $\Psi_{\rm E}$ is by incorporating corrections, term by term, to $\tilde{a}_{\rm E}$ that cancel the contribution coming from $\tilde{a}_j$ in Eq. \eqref{aEc} up to a higher order. However, this method is even more involved than the one we have proposed along this section.

\subsection{IV.\quad Robustness of the massive Dirac edge states to disorder}
\label{app_D}
In this Section we show that the massive Dirac states are robust against the presence of a disordered potential. To this end, we add a random chemical potential 
\begin{equation}\label{ham_disorder}
H_{\delta_\mu} = \sum_i \epsilon_i a^\dag_i a_i
\end{equation}
to the Hamiltonian (1) of the main text with only long-range pairing terms. The coefficients $\epsilon_i \in [-\delta_\mu,\delta_\mu]$ are chosen from a random uniform distribution  with zero mean value and width $2\delta_\mu$. 

\begin{figure}
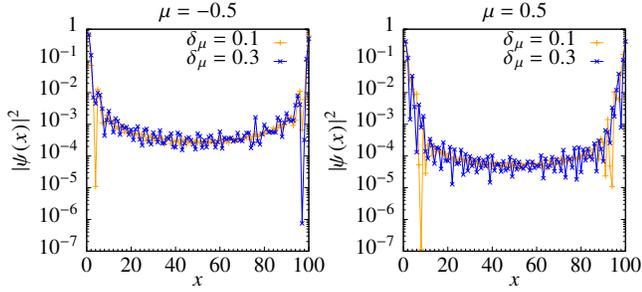

\includegraphics[scale=0.31]{{{wavefunction_mu_-0.5}}}%
\includegraphics[scale=0.31]{{{wavefunction_mu_0.5}}}%
\caption{Spatial probability distribution $|\psi(x)|^2$ for the massive Dirac edge mode for different chemical potentials $\mu$ and disorder strengths $\delta_\mu$. The wave functions are localised at the ends of the chain, even in the presence of a disordered potential.}\label{fig_disorderWF}
\end{figure}

Figure \ref{fig_disorder} shows the energy spectrum $E_n$ (averaged over 100 disorder realizations) of the total Hamiltonian $H+H_{\delta_\mu}$ for $\alpha=0.5$ in the massive Dirac sector for two values of $\delta_\mu$.  It is possible to see that, when $\mu<1$, one state (plotted in green) is separated by a finite gap from the band of the bulk states (plotted in red). This state is still an edge state as Fig.~\ref{fig_disorderWF} shows. There we plot the spatial distribution  $|\psi(x)|^2$  of the wave function of the mode lying outside the band of the bulk states for two values of $\delta_\mu$ and $\mu$ for a system of $L=100$ sites. Even in the presence of a random potential term, this state is a Dirac massive edge state as it is always localised at the ends of the chain.

\subsection{V.\quad Topological Quantum Memory}
\label{app_E}

As stated in the main text, it is possible to define a topological qubit using the new non-local massive Dirac fermions. In the short-range Kitaev chain, the topological protection of the unpaired Majoranas is related to the conservation of fermion parity and the gap isolating the MZMs from the bulk states. We want to stress that these same features also hold true for non-local massive Dirac fermions. 

\begin{figure}[t]
\vspace{0.3cm}
\centering
\includegraphics[width=\columnwidth]{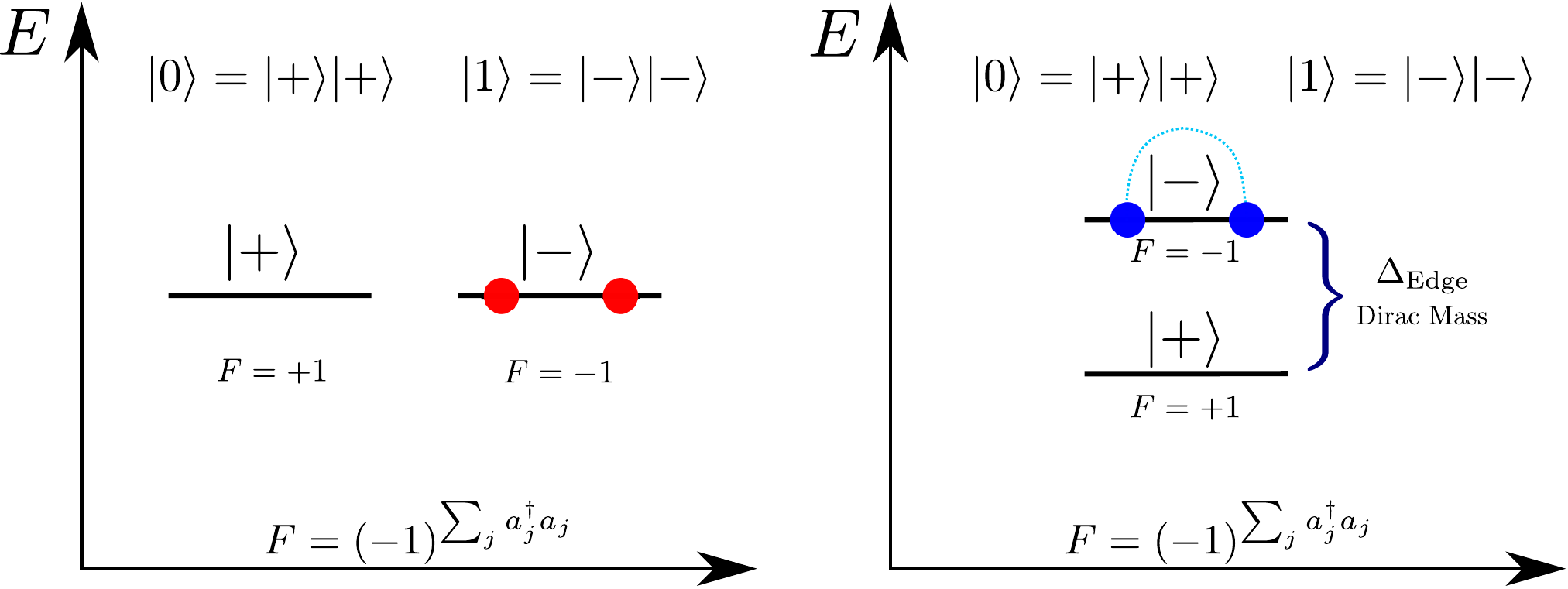}
\caption{At the l.h.s. we have depicted the construction of a qubit within the even parity sector for the short-range case. The two level system $\ket{+},\ket{-}$ represent whether the two MZMs at the edge are populated or not. At the r.h.s. we can see a similar scheme for a qubit in the even parity sector within the Dirac phase. The only difference comes at level of the states $\ket{+},\ket{-}$, representing whether the non-local massive Dirac fermion is populated or not.}
\label{TQC_Dirac}
\end{figure} 

As depicted in Fig. \ref{TQC_Dirac}, we can still define even and odd parity states $\{ \ket{+}, \ket{-} \}$, respectively. These two states have different fermion parity $F$, physically depending on whether we populate the non-local Dirac fermion or not:
\begin{equation}
\tilde{a}_{\rm E}\ket{+}=0,~~~~\ket{-}=\tilde{a}^{\dagger}_{\rm E}\ket{+}.
\end{equation}

The fact that the effective two level system is gapped is irrelevant in that respect. Additionally, as we show in Fig.~\ref{fig_disorder}, this effective two level system is separated from the bulk eigenstates by an energy gap and it is robust against disorder perturbations. Using these states, we can define a qubit using two copies of the Kitaev chain. The reason behind is the impossibility to have a qubit without a definite fermion parity \cite{rmp3,rnpj1}. Therefore, the qubit can be defined either in the even or odd parity sector. As shown in Fig. \ref{TQC_Dirac} of the SM, we define a qubit with even fermion parity, as $\ket{0}=\ket{+}\ket{+}$ and $\ket{1}=\ket{-}\ket{-}$.

Furthermore, proposals to perform topological quantum gates with Majorana fermions based on their braiding properties have been recently proposed \cite{Alicea_et_al_11,Baranov_et_al_13}. All the physical operations needed can be written in terms of fermionic degrees of freedom, involving on-and-off switchings of the different coupling constants. For the present case with long-range couplings, a more detailed analysis in order to elaborate a concrete proposal would be required. This is out of the scope of the present work but it is left as an outlook of the paper.

\end{document}